\documentclass[aps,prl,twocolumn,tightenlines,amsmath,amssymb,superscriptaddress]{revtex4}

\usepackage{graphicx}
\usepackage{amsmath}
\usepackage{amssymb}
\usepackage{epstopdf}
\usepackage{color}
\usepackage{lipsum}
\usepackage{ulem}
\usepackage[dvipsnames]{xcolor}

\DeclareGraphicsExtensions{.eps}

\newcommand*{\rom}[1]{\expandafter\@slowromancap\romannumeral #1@}

\begin{document}

%\title{\nb{ Solid-state} spin dynamics and coherence via  acoustic phonon-assisted excitation \nb{ preserving polarization selection rules} }
%\title{Probing a confined hole spin dynamics and coherence via acoustic phonon-assisted excitation  }

\title{Probing the dynamics and coherence of a semiconductor hole spin via acoustic phonon-assisted excitation}

\author{N. Coste}
\author{M. Gundin}
\author{D. Fioretto}
\author{S. E. Thomas}
\author{C. Millet}
\author{E. Medhi}
\author{M. Gundin}
\affiliation{Université Paris-Saclay, CNRS, Centre de Nanosciences et de Nanotechnologies, 91120, Palaiseau, France}
\author{N. Somaschi}
\affiliation{Quandela SAS, 10 Boulevard Thomas Gobert, 91120, Palaiseau, France}
\author{M. Morassi}
\author{M. Pont}
\author{A. Lema\^itre}
\author{N. Belabas}
\author{O. Krebs}
\affiliation{Université Paris-Saclay, CNRS, Centre de Nanosciences et de Nanotechnologies, 91120, Palaiseau, France}
\author{L. Lanco}
\affiliation{Université Paris-Saclay, CNRS, Centre de Nanosciences et de Nanotechnologies, 91120, Palaiseau, France}
\affiliation{Universit\'e Paris Cit\'e, Centre for Nanoscience and Nanotechnology (C2N), F-91120 Palaiseau, France}
\author{P. Senellart}
\affiliation{Université Paris-Saclay, CNRS, Centre de Nanosciences et de Nanotechnologies, 91120, Palaiseau, France}

%\date{August 2021}

	\begin{abstract}
		 Spins in semiconductor quantum dots are promising local quantum memories to generate polarization-encoded photonic cluster states, as proposed in the pioneering Rudolph-Lindner scheme ~\cite{lindner_proposal_2009}. However, {harnessing} the polarization degree of freedom of the optical transitions is  {hindered by} resonant excitation schemes  that {are}  widely used to obtain high photon indistinguishability. Here we show that  acoustic phonon-assisted excitation, a scheme that preserves high indistinguishability, also allows to fully exploit the polarization selective optical transitions to initialise and measure single spin states. We access the coherence  of  hole spin systems in a low transverse magnetic field and directly monitor the spin Larmor precession both during the radiative emission process of an excited state or in the quantum dot ground state. We report  a spin state detection fidelity of $94.7 \pm 0.2 \%$ granted by the optical selection rules and a $20\pm5$~ns hole spin coherence time,  {demonstrating}the potential of this scheme and system to generate  linear cluster states with a dozen of photons. \end{abstract}

\maketitle

%section{Introduction}
%\green{We need to introduce a bit more smoothly the concepts of the first sentence.}

The spins of carriers in solid-state systems are of great interest for optical quantum technologies~\cite{gao_coherent_2015}, be it for distant quantum node entanglement~\cite{delteil_generation_2016,Stockill2017},  envisioned photon-photon gates~\cite{Bonato2010}  or multi-photon entanglement~\cite{lindner_proposal_2009,schwartz_deterministic_2016}. %QD possibilities can be enhanced as they can confine a single carrier, electron or hole, by engineering of the sample structure ~\cite{hilaire_deterministic_2020}.
%In this case, spin-dependant polarization selection rules arise and can be used to generate entanglement between the spin of the confined carrier and emitted photons. This fundamental feature of charged quantum dots has led to the development of spin photon interface ~\cite{yilmaz_quantum-dot-spin_2010, ding_coherent_2019, appel_coherent_2021}, photon-to-spin teleportation ~\cite{gao_quantum_2013}, remote spin entanglement ~\cite{delteil_generation_2016, stockill_phase-tuned_2017}, and is currently investigated in the perspective of multi-photon entangled states generation ~\cite{schwartz_deterministic_2016}.
In 2009, Lindner and Rudolph (LR)~\cite{lindner_proposal_2009} proposed a scheme which harnesses the spin properties of semiconductor quantum dots (QDs) to tackle the great challenge of deterministic photon cluster state generation. These multi-entangled photon  states are highly sought-after   to implement all-optical quantum networks~\cite{Zwerger2012,Azuma2015} as well as measurement-based quantum computing~\cite{raussendorf_measurement-based_2003}. The LR scheme considers a {periodically excited} quantum dot electron spin precessing around a weak magnetic field: by {synchronizing} the excitation pulses and the spin precession period, a string of polarization-encoded single photons can be generated, forming a one-dimensional cluster state.  The scheme is conceptually simple, robust and relies on polarization selection rules and a spin coherence time much longer than the radiative lifetime of the optical transitions. %Moreover,  one can obtain higher dimensionality cluster states by making use of fusion gates if the generated photons are indistinguishable ~\cite{Rudolph2013}. 
%{Moreover, \PS{ high photon collection can be obtained inserting the QD in cavities~\cite{somaschi_near-optimal_2016,Wang2016} and} indistinguishable generated photons give access to higher dimensionality cluster states~\cite{Rudolph2017}, instrumental for error correction quantum computing, via fusion gates.}
In order to obtain high multi-photon generation rate, a high photon collection efficiency is required which can be obtained by inserting the QD in cavities~\cite{somaschi_near-optimal_2016,Wang2016}. Furthermore, if the photons are highly indistinguishable then higher-dimensional cluster states can be generated via fusion gates~\cite{Rudolph2017}, which are instrumental for error-corrected quantum computing

A first implementation of the LR scheme was demonstrated using a spin system consisting of a dark QD exciton ~\cite{schwartz_deterministic_2016}. However, the use of non-resonant excitation limited the photon indistinguishability to  $\approx 17\%$~\cite{Cogan2021}. More recently, a similar demonstration was demonstrated using a QD hole spin with a non-resonant scheme exploiting the fast emission of longitudinal optical phonons~\cite{Cogan2021}. 
The use of optical-phonon relaxation allows for single-photon indistinguishabilities of over $90\%$ for QDs in bulk, which typically have a nanosecond radiative lifetime. However,  the intrinsic time jitter of the LO phonon relaxation becomes comparable to the optical transition lifetime when the QD is inserted in a cavity and the photons indistinguishability is reduced~\cite{Kiraz2004}.

%Such scheme  allows  indistinguishabilities above $90\%$ for QDs in bulk showing  nanosecond typical lifetimes. However, for QDs in cavities, this value is expected to decrease as the intrinsic time jitter of the phonon relaxation becomes comparable to the Purcell-enhanced optical transition lifetime.}% but not for QDs in cavities  because of the intrinsic time jitter of the phonon relaxation that becomes comparable to the optical transition lifetime~\cite{Kiraz2004}.} %- % and high polarization purity, high photon collection efficiencies, and a spin coherence time much longer than the radiative lifetime of the optical transition. The aim of this work is to benchmark relevant parameters for implementing polarization-based cluster states generation schemes.
 
The generation of highly indistinguishable photons has been reproducibly demonstrated through resonant excitation of QDs~\cite{He2013,Senellart2017}, and with unparalleled efficiency when making use of microcavities~\cite{somaschi_near-optimal_2016, Wang2016, Tomm2021}. However, resonant excitation schemes in a confocal microscopy require rejection of the resonant excitation laser through polarization filtering, which precludes polarization encoding of single photons {and hence} using the LR scheme. %Polarized cavities and resonant excitation thus rule out the possibility to implement the Lindner and Rudolph scheme. So far, this protocol has only been implemented using a dark exciton state as an entangler ~\cite{schwartz_deterministic_2016}, however this approach is limited by xxx. 
Alternative schemes have been proposed to address this difficulty, replacing the polarization degree of freedom by time bin encoding ~\cite{Tiurev2021,Tiurev2022,Lee2019}. This is however more resource demanding as it involves an intense magnetic field and phase stabilization of the emitted photons. % and alignment of the cavity axis with the {magnetic} field. % B /cavity mode orientation, might need ultranarrow cavity, or broadband + polarized, or to
%It was recently demonstrated \PS{both theoretically and experimentally} that a longitudinal acoustic (LA)-phonon assisted excitation scheme allows the generation of single photons with high  indistinguishability as well as high occupation probability {of the QD excited states after excitation,} enabling  high brightness. 
%To circumvent this spin/optimized SPS compatibility bottleneck, 
{Longitudinal acoustic (LA)-phonon assisted excitation schemes have been proposed for the generation of single photons with near-unity  indistinguishability as well as high occupation probability of the QD excited states~\cite{Finley2014,Cosacchi2019,GustinHughes2020}. In this scheme, phonon-assisted processes do not lead to time jitter of the photon emission since the phonon relaxation only takes place while the excitation pulse is on. This scheme hence allowed the demonstration of high indistinguishability and high efficiency single photon emission for QDs in microcavities~\cite{thomas_bright_2021}.}

In the present work, we {experimentally} demonstrate that acoustic phonon-assisted excitation  allows to fully exploit the polarization-selective optical selection rules of  QDs charged by a single carrier (electron or hole). We demonstrate that %it allows
{this excitation scheme is compatible with subsequent} monitoring {of} the dynamics and coherence of a single hole spin. We {first} observe the precession of a hole spin in a low transverse magnetic field through polarization and time resolved photoluminescence of a negatively charged QD under pulsed excitation. These measurements also allow to extract a lower bound to the spin coherence time limited by the %short 
{emission} lifetime of the excited state. To access longer spin coherence times, we perform polarization-resolved 2-photon correlation measurements of a positively charged QD under continuous wave phonon-assisted excitation. Our measurements demonstrate a hole spin coherence time  exceeding the radiative lifetime by more than one order of magnitude, an important step towards the implementation of the original Lindner and Rudolph scheme.% \nb{Polarization purity of the emitted photons \PS{above 95\%} which is a further step forward. }

%sto probe the polarization of the photon emitted by charged excitons.  The LA-phonon excitation is fully compatible with an unpolarized cavity, which enhances the QD emission~\cite{thomas_bright_2021} while preserving an high polarization purity. %These cavities
%Moreover, spectral filtering of the excitation laser allows to excite and collect in any given polarization with high collection efficiencies thus enabling polarization encoding schemes. 
%In the scheme we demonstrated, we benefit from an unpolarized Purcell-enhanced radiative lifetime : this ....larger states  Lindner and Rudolph proposal implementation. 
%Here, we use LA excitation on charged excitons in InGaAs quantum dots and extract all the relevant parameters on the dynamic of the QD system to implement the LR scheme. 
 
%While significant progress has been done towards both optimized single photon generation as well as optimal control of the spin ~\cite{hanson_coherent_2008, press_complete_2008, greilich_ultrafast_2009, godden_coherent_2012, gao_coherent_2015}, combining the two aspects remains challenging. Indeed, achieving efficient QD-based single photon sources using highly polarized cavities results in sacrificing spin selective transitions. The cavity indeed favors the emission of linearly polarized light and thus erases the link between spin and photon polarization which is instrumental in the LR protocol.

We study annealed InGaAs quantum dots emitting around {$\lambda$=}927 nm, embedded in a planar GaAs $2{\lambda}$-cavity, surrounded by 16 pairs of AlAs/GaAs distributed Bragg reflector (DBR) on top and 34 at the bottom. The cavity region starts with a $\frac{\lambda}{2}$ GaAs layer below the quantum dots, with a n-type doping at $10^{18}~$cm$^{-3}$ stopping $25$ nm before the QD layer. Above the QD layer, a $8$-nm GaAs capping layer is followed by %a
{an undoped} Al$_{0.33}$Ga$_{0.7}$As superlattice~\cite{Tomm2021}, with a total optical %distance
{thickness} of $\frac{3\lambda}{2}$. %undoped.
In the Bragg mirrors, the n-type (bottom) and p-type doping (top) are gradually increased {further from the cavity} to ensure good ohmic contacts: %application of a 
{a voltage} bias then allows  a control of the charge  of the quantum dot ground state. 

The LA phonon-assisted excitation scheme is sketched in Fig.1 a. The {pulsed excitation} laser is blue detuned from the optical transition of the QD by roughly $\Delta \lambda =$ 1 nm. During the  excitation pulse (here $\approx 20\ $ps), the laser  adiabatically dresses and undresses the QD ground and excited states while a fast relaxation of a longitudinal acoustical phonon allows for an efficient excited state population~\cite{Finley2014}. Overall, this %realizes
{achieves} an incoherent population inversion, leaving the QD in the excited state with a  probability above 85\%~\cite{thomas_bright_2021}.  In the present work, we also implement a continuous-wave (c.w.) LA phonon-assisted excitation scheme, that, while not allowing for high occupation probability, enables %exploring
{the exploration of} long delay spin dynamics. The single photon emission from the QD is spectrally filtered from the excitation laser using high transmission, narrow bandpass filters (0.8 nm bandwidth) with no need for polarization filtering. %This allows for excitation and collection of single photons with arbitrary polarization~\cite{thomas_bright_2021}. 
The QD emission is then analyzed in polarization with waveplates and a polarizing beam-splitter, and detected with fast superconducting nanowire single photon detectors (time jitter $\approx 32\ $ps).

\begin{figure}
	\includegraphics[width=\linewidth]{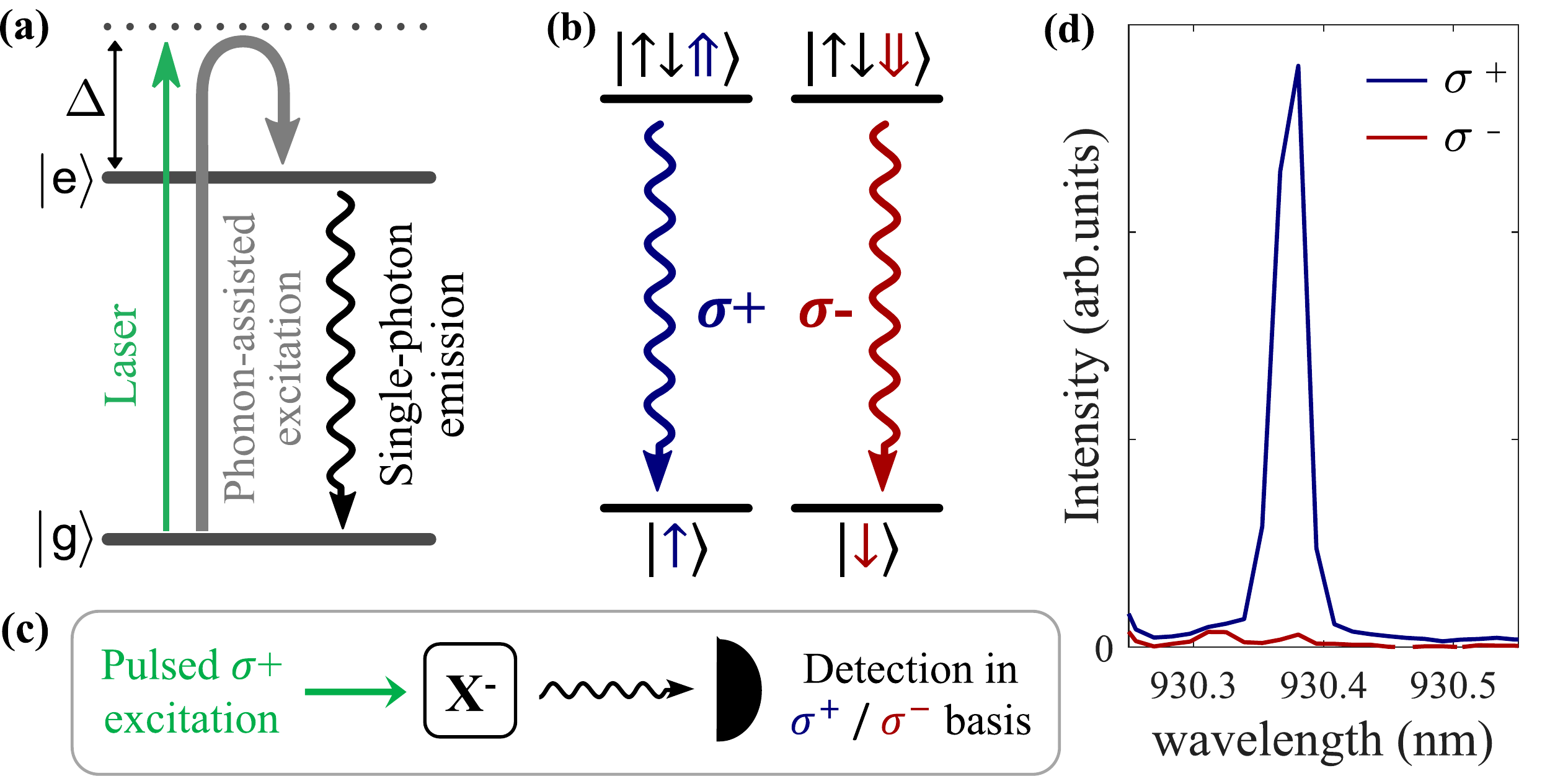}
	\caption{ { (a) Schematic of the principle of LA excitation where a 20 ps laser pulse is detuned by $\Delta\lambda$ = 1 nm from the trion transition. Acoustic phonon-assisted relaxation  takes place between the laser-dressed QD states. Adiabatic undressing of the laser leaves the quantum dot in an excited state with high probability. (b) Energy levels and optical selection rules of the negatively charged QD in absence of magnetic field, with spin projections along the QD growth axis. (c) Schematic of the {photoluminescence} experiment. The negatively charged quantum dot is excited with $\sigma^{+}$
	 polarized laser pulses. The single photon emission is detected in the circular polarization basis {using waveplates and polarizers}. (d) Time-integrated  spectrum of a negative trion under pulsed, circularly polarized acoustic phonon assisted-excitation.
	} \label{Fig_1}}
\end{figure}

 We first study the properties of the hole spin in the excited state of a negatively  charged QD. The  ground state consists of two degenerate electron spin states in the absence of magnetic field. In the excited state, an electron-heavy-hole pair is optically created in addition to the resident electron ($X^-$ trion state). The two electron spins form a spin singlet, and  the magnetic properties of the trion are determined by those of the unpaired hole only. % the contribution of the hole spin remains. % becomes the spin of the excited state. %(mention heavy/light hole, or too much details?) 
Due to orbital angular momentum conservation, ground and excited states spin projections along the growth axis of the quantum dots ($\pm 1/2$ for the electron, $\pm 3/2$ for the heavy hole) result in polarization selective optical transitions involving circularly polarized photons ($\sigma ^{\pm}$). 
%\red{The arrows between ground (excited) states illustrate} the difference between the electron spin (hole spin) that arises from the wave function of the carrier : electron spins have a s-type wave function, overlapping with the nuclei of the quantum dot over $\sim 10^5$ lattice sites, while hole spins have a p-type wave function whose antinodes coincide with the lattice sites~\cite{prechtel_decoupling_2016}. The electron spin is therefore  more sensitive to hyperfine interaction with the nuclei, leading to rapid decoherence whereas the hole spin are less sensitive to the fluctuating nuclear spin environment. 
The {excitation} is circularly polarized $\sigma^+$ pulsed laser light and the  power is chosen to maximize the QD single-photon emission. Following the optical selection rules, the circularly polarized excitation populates only the trion excited state $|\uparrow \downarrow \Uparrow \rangle$ {(Fig. 1b)}. The single-photon emission is measured either along a parallel $\sigma^+$ or %perpendicular
{orthogonal} $\sigma^-$ circular polarization {(Fig. 1c)}. If the  hole remains in the same excited state during the whole spontaneous emission process,  the single-photon is emitted with a polarization parallel to {that of} the laser. Emission in the orthogonal polarization can only occur if the QD undergoes a spin flip in the excited state within the spontaneous radiative decay lifetime.  The time-integrated  spectra in figure 1.{d} show hardly any emission  in the polarization orthogonal to the excitation so that the corresponding time and spectrally-integrated intensities $I_{\sigma^-}$ and $I_{\sigma^+}$ verify {$I_{\sigma^-} << I_{\sigma^+}$. This corresponds} to a time-integrated degree of circular polarization, $DCP = \left( \frac{I_\mathrm{\sigma_\mathrm{+}} - I_\mathrm{\sigma_\mathrm{-}}}{I_\mathrm{\sigma_\mathrm{+}} + I_\mathrm{\sigma_\mathrm{-}}} \right)$, of $94.7 \pm 0.2 \%$ and demonstrates the excellent preservation of the polarization selection rules when using phonon-assisted excitation processes in contrast  to other incoherent excitation schemes such p-shell excitation. This is consistent with previous observations for neutral excitons in both GaAs~\cite{Rastelli2019} and InGaAs QDs~\cite{thomas_bright_2021}.%, his confirms the observation of (ref rastelli paper) \nb{made in the context of ...} that LA excitation highly preserves the polarization selection rules, particularly compared %NB: il faut quelques mots de plus pour se différencier de ce papier}  In addition, this value can be compared to the DCP of positively charged QD of  ??? \% (data not shown) that confirms that hole spins are less sensitive than electron spins to depolarizing hyperfine interaction with the surrounding nuclear spins \cite{REFS: merkulov, warburton...}. In this work, we focus on negatively charged QD and hole spins as they are thus the best candidates for LR cluster states generation.} 
The same excitation scheme allows {for} monitoring the coherent Larmor precession of the hole spin in the excited state around an applied in-plane magnetic field (Voigt configuration). The magnetic field lifts the degeneracy of both ground and excited states resulting in an energy splitting $\Delta E_\mathrm{e/h}$ given by $\Delta E_\mathrm{e/h} = \hbar g_\mathrm{e/h} \mu_\mathrm{B} B$  where $g_\mathrm{e/h}$ {are} the transverse electron {or} hole Land\'e factor, $\mu_B$ is the Bohr magneton and B is the magnetic field strength. In this context, the states in Fig. 1.b are no longer eigenstates of the system Hamiltonian.
At time $t_0$, the QD is excited using a short $\sigma^+$-polarized  pulse laser pulse (i.e. of bandwidth greater than  $\Delta E_\mathrm{e/h} / \hbar$), and is initialized in state $|\uparrow \downarrow \Uparrow \rangle$ (depicted as $|\Uparrow \rangle$ in Fig. 2.a. for simplicity). The hole undergoes Larmor precession due to the magnetic field, and periodically oscillates between states {$| \Uparrow \rangle$ and $| \Downarrow \rangle$.} At a later time $t_0+\tau$, the QD  decays and emits either a $\sigma^+$ or $\sigma^-$ polarized photon, depending on the spin state after the precession time of $\tau$.

% As illustrated in Fig. 2a, the preparation of state $|\uparrow \downarrow \Uparrow \rangle$ { -- pictured  $|\Uparrow \rangle$ in the excited state for simplicity --} through the  excitation by a $\sigma^+$-polarized laser pulse, at $t_0$ triggers a Larmor precession between states %$|\uparrow \downarrow \Uparrow \rangle$ and $|\downarrow \uparrow \Downarrow \rangle$. {$| \Uparrow \rangle$ and $| \Downarrow \rangle$.} Consequently, trion decay at a later time $t_0+\tau$ can lead to the detection of a $\sigma^+$ or $\sigma^-$ polarized photon, depending on the trion state.

Figure 2.b shows the signature of the Larmor precession through the temporal evolution of the emitted intensities $I_\mathrm{\sigma_\mathrm{+}}(t)$ and $I_\mathrm{\sigma_\mathrm{-}}(t)$ during trion decay, for increasing applied magnetic fields up to 450 mT. Periodic oscillations are observed as the hole spin in the trion state coherently precesses at the Larmor frequency $g_\mathrm{h} \mu_\mathrm{B} B$.   %As expected, we observe a linear dependency on the beatings frequency as shown in Fig. 2.d. from which we retrieve the hole transverse g-factor \blue{ $g_h = 0.38\pm 0.04$, a value consistent with previous measurements on annealed InGaAs QDs \cite{Trifonov2021}}. \red{Please note that the Landé factor is 0.38, not 0.19 as stated before. Correct this factor 2 in the future version of the figures. NC : got it, it came from the fact that I was fitting the beatings with a cos² function, and cos²(x) = (cos(2x)+1)/2}
The sum of the two intensities (black line) is also displayed in the top panel of Fig. 2.b, displaying a monoexponential decay associated with the trion lifetime. Figure 2.c. presents the time-resolved {$DCP$}, which demonstrates only a slight damping of the oscillations over time. This indicates that the spin maintains its coherence well beyond the timescale of the radiative decay process.%, i.e. the intensity contrast, whose oscillation preserves most of its coherence during the radiative process. % giving a lower bound on the hole spin coherence time of several ns.
%Finally, Fig. 2.d. displays the frequency of the Larmor precession as a function of the magnetic field intensity, showing a linear dependence.

	\begin{figure}[t]
	\includegraphics[width=0.9\linewidth]{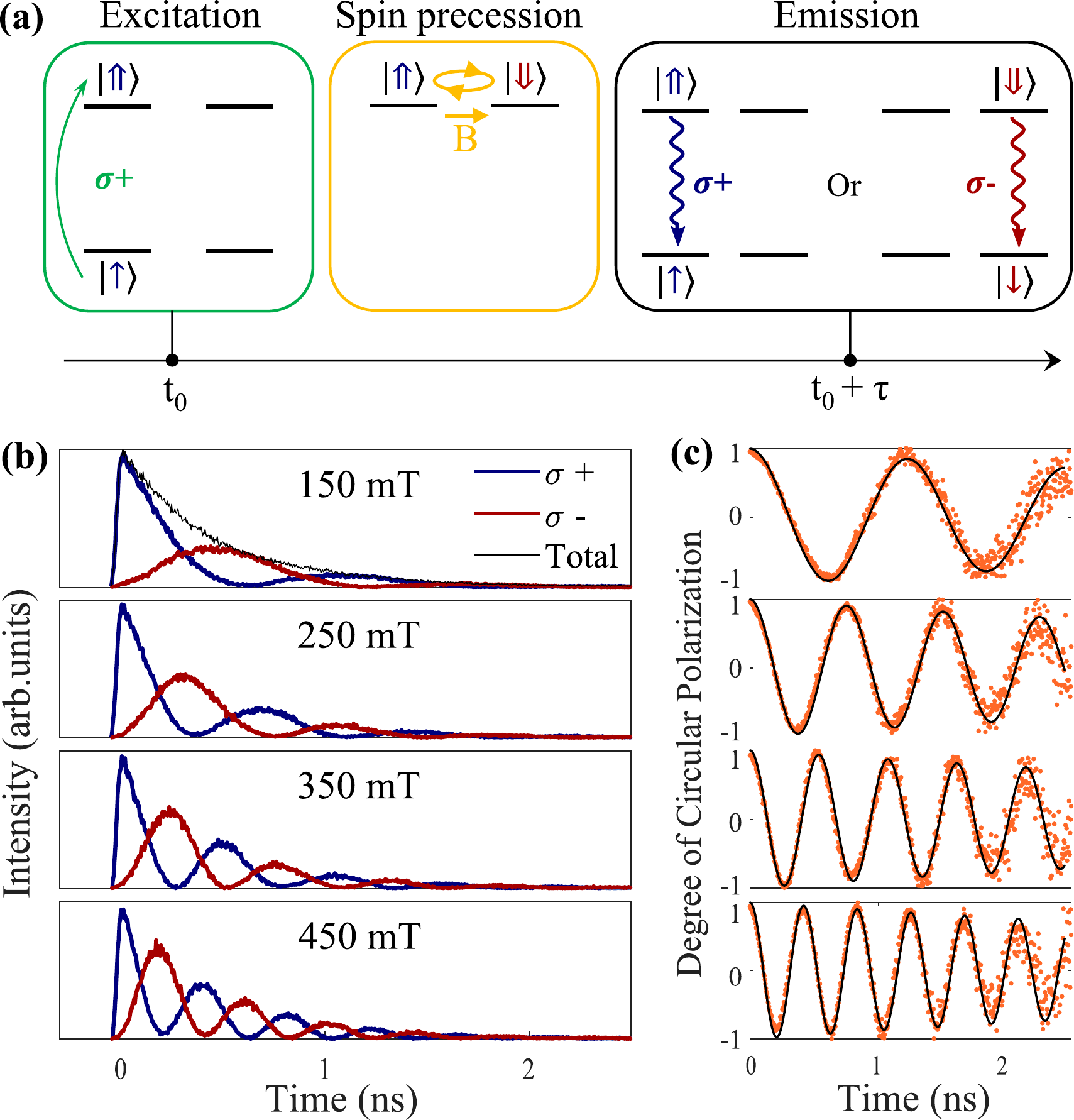}
	\caption{ (a) Schematic of the
	% experiment 
	{excited-state spin dynamics} measurement. The negatively charged quantum dot is excited with $\sigma^+$ polarized laser pulses, populating only the state $|\uparrow \downarrow \Uparrow \rangle=|\Uparrow \rangle$. The in-plane magnetic field $B$ induces a Larmor precession between the two trion spin states, and the emitted photons are detected in the circular parallel $\sigma^+$ or perpendicular $\sigma^-$ polarization. (b) Polarization resolved time trace of the QD emission under in-plane magnetic field, and (c) corresponding degree of circular polarization $DCP$. The beatings reflect the coherent Larmor precession of the hole spin around the magnetic field. The  lines are fit to the experimental data (see text). %\PS{Please Nathan remove figure d - maybe you can put figure b and c next to each others.} %\green{(LL: the following part should be removed) NB: why ?} (d) Zeeman splitting as a function of the magnetic field intensity, deduced from a fit of the previous curves. The slopes gives an out-of plane Landé factor of g=0.38 \label{Fig_PrecessionExc}
	}
\end{figure}
We fit the evolution in Fig 2c using a simple model, where the  master equation is numerically solved
%, in presence of several simplifications.
considering  that the LA-assisted excitation process results in the same trion state populations as for resonant excitation. We also assume that the detected intensity $I_\mathrm{\sigma_\mathrm{+}}$  (resp. $I_\mathrm{\sigma_\mathrm{-}}$) is directly proportional to the population of the trion state $|\uparrow \downarrow \Uparrow \rangle$ (resp. $|\downarrow \uparrow \Downarrow \rangle$) at any given time and we neglect the hyperfine interaction on the hole spin in the trion state. These measurements give access to an effective hole spin coherence time $T_2^*$ which is dominated by a pure dephasing term. In the master equation of the model $T_2^*$ is then reduced to a homogenous pure dephasing time. %Since these measurements give access to a total spin coherence time, we model the spin decoherence by an empiric coherence time $T_2^*$ corresponding to a pure dephasing term.%and spin-flip mechanisms. Here,  we arbitrarily choose to include them in the form of a pure dephasing term. %Note that  %Identical results are  obtained by including a spin-flip term: the only accessible quantity is the total spin coherence time, that can derive both from pure dephasing and spin-flip mechanisms."

With these hypothesis, the master equation yields a good fit %allows obtaining a satisfying fit
for all the spin evolutions in Fig. 2c.  {using a trion radiative decay time $T_1^{\mathrm{(trion)}}=450 \pm 20$ ps, $T_2^*\ge15 \pm 5$ ns, and  $g_h=0.38 \pm 0.01$. } This measurement provides a lower  bound for  $T_2^{\mathrm{(spin)}}$, limited by the lifetime of the excited state.
The relatively high value of the transverse hole spin Land\'e factor $g_h$, which is typical in strongly annealed quantum dots \cite{Trifonov2021}, also implies that a relatively weak magnetic field is sufficient to implement a large number of precession cycles during the spin coherence time, which is significantly longer than the trion lifetime. All conditions are thus met to implement the LR scheme for  cluster state generation with a dozen of photons.

\begin{figure*}[t]
	\includegraphics[width=0.9\linewidth]{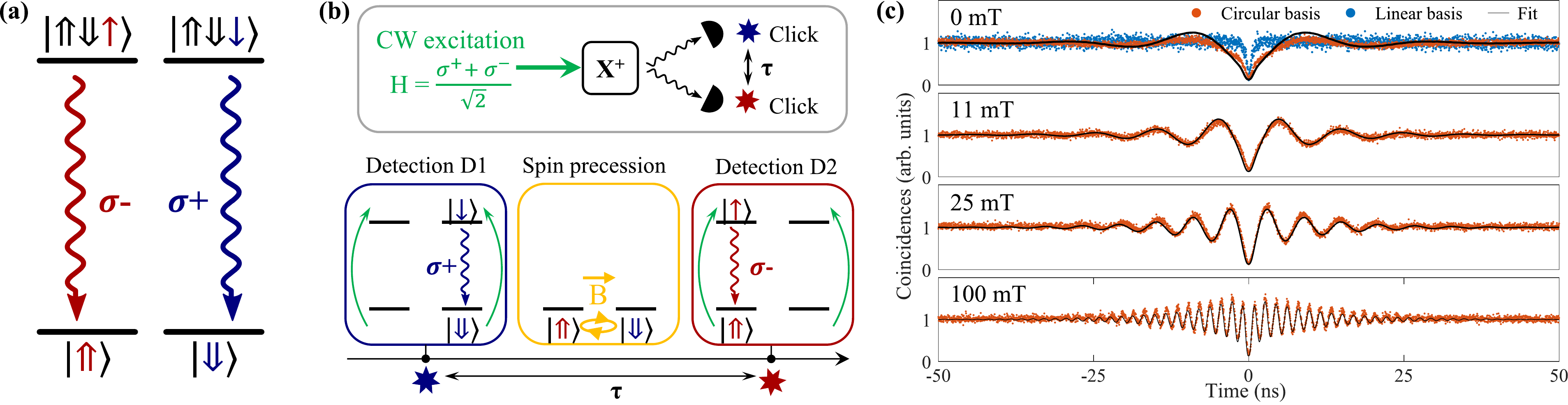}
	\caption{  (a) Energy levels and optical selection rules of the positively charged exciton with spin projections along the QD growth axis. (b) Schematic of the experiment. The positively charged quantum dot is LA-phonon assisted excited with linearly polarized, continuous laser light. The single photon emission is projected along two orthogonal bases with a set of waveplates and a polarizing beamsplitter. Cross-correlations are measured between the two detectors. (c) Cross-correlation measurements between orthogonal polarizations in the circular basis ($\sigma_\mathrm{+}$/$\sigma_\mathrm{-}$) (orange), and a linear basis (blue).  The lines are fit to the experimental data (see text).%The oscillations in the $\sigma^+$/$\sigma^-$ basis reflect the precession of the hole spin in the ground state. The damping of the oscillations provide a lower bound on the hole spin coherence time. When measuring in a linear polarization basis, no spin-dependant correlation can be observed, and the dip at 0 delay is characteristic of the single-photon nature of the emission.
	\label{Fig_PrecessionCW}}
	%Hole spin (ground state) precession of a positive trion. the positive trion is excited in LA CW. Its single photon emission is analyzed in the R/L basis... correlations RL, a click in a detector projects and heralds the spin state,  (b) fit of the precession frequency vs B (do we need this plot? Redundant with fig.2c), and (c) fit of the exponentially decaying enveloppe of the correlations as a function of the magetic field, corresponding to a lower bound on the hole spin coherence time.(put B values)
	%Nadia proposal: Dynamic of the ground state hole spin precession  of a single positive trion X$^+$ : The positive trion is excited in LA CW Start-Stop configuration (see text) . Its single photon emission is analyzed in the sigma + / sigma  basis by means of scheme (a) that measures correlations :  a click in a detector corresponds to the hole spin projection and the start of the precession. ??? It heralds the spin state, (b) ?? fit of the precession frequency as a function of in plane B and matching Fig 2c data, and (c) Exponential decay fitted from the enveloppes of the correlations in (b) as a function of the magnetic field intensity,  giving access to a lower bound of the ground state hole spin coherence time (at B =0, 11, 25 100 mT)} \label{Fig_PrecessionCW}}
\end{figure*}

{The polarisation-resolved lifetime measurements thus give direct access to the properties of a hole spin in the excited state, but only provide a lower bound to the spin coherence time when it exceeds  the trion lifetime by orders of magnitude. Moreover, the LR scheme would {be more readily implemented with} %favourably exploit 
a hole spin in the ground state to generate many-photon linear cluster states owing to the reduced decoherence induced by nuclear spins~\cite{Urbaszek2013}. To access the hole spin coherence time in the ground state, we %\green{now}
propose and implement correlation measurements under continuous wave excitation. 

We now consider a positively-charged QD with a hole in the ground state, and  and excite the positive trion transition again using the phonon-assisted excitation scheme. Figure 3.a. presents the  optical selection rules corresponding to this situation, % where the spin decoherence induced by the nuclei now affects mostly the excited state 
{while Figure 3.b. details the experimental scheme used to probe the ground state hole spin dynamics.
%presents the new experimental conditions.
T}he excitation laser is now linearly polarized. In the absence of spin initialisation, the hole spin is in a mixed state with equal contributions from the $| \Uparrow \rangle$ and $| \Downarrow \rangle$ states, so that both trion transitions can be excited. The {hole spin dynamics is then probed by recording the statistics of circularly cross-polarized emitted photons i.e. counting coincidence events between 
a {$\sigma^+$}-polarized detected photon, and another {$\sigma^-$}-polarized one, separated by a delay $\tau$. {Indeed, w}hen a first photon is detected at time $t_0$ in polarization {$\sigma^+$}, this leads to a projective spin measurement in the $|\Downarrow \rangle$ ground state. This is followed by a precession of the ground state hole spin around the applied transverse magnetic field. 
The subsequent detection of a second $\sigma^-$-polarized photon at time $t_0+\tau$ will therefore reflect the  Larmor oscillation of the hole spin in the QD ground state after its initialization. 
We can thus measure the precession of the hole spin in the ground state through polarization-resolved cross-correlations between the two detectors. Mapping those cross-correlations to the spin population relies on using a low enough excitation power  to limit multiple re-excitations. We also note that exciting with linear polarization prevents  a dynamic nuclear polarization - i.e. polarisation of a  nuclear spin induced by electron-nuclei spin flip-flop processes in the excited state~\cite{Urbaszek2013}. Finally, the $X^+$ trion spin precession in the applied field or random Overhauser field during the trion radiative lifetime, is now time-averaged, and therefore reduces the contrast of the measured  oscillations. 

%\green{LL: au delà de l'explication des polars D et A, j'ai aussi un souci avec le 0T, qui n'est pas vraiment à 0T (Manuel a fitté avec un décalage de 5-7mT je crois, pour toutes les figures). Du coup il y a une ondulation que le texte ne commente pas, mais qu'un referee pourrait repérer. Je crois que Manuel a plein de fits disponibles et on pourrait commencer sans le 0T, je pars sur ça pour la rédaction, et mets en commentaire la version précédente. NB: j'ai ébauché 2 phrases rapides *** pour réconcilier le tout. Il me semble que c'est faisable...}

{Fig. 3.c. shows the measured oscillations for increasing magnetic fields. T}he cross-correlations are measured through the coincidences between $\sigma^+$ and $\sigma^-$ detection events, normalized to unity at long delays. At very short delays, an anti-correlation is always observed due to the single-photon nature of the emitted field. % : after projection in one of the ground states, the trion state remains empty for a few hundred picoseconds, until it is repopulated by the CW excitation, and \nb{emit} a second photon. 
{This  antibunching is visualized when using a linear polarisation detection basis to remove the signature of spin-dependent correlations (blue data in the top panel of Fig. 3.c.). In the circular polarisation basis, a}fter $1-2$ ns, we observe a decaying oscillation of the cross-correlation signal, which is a signature of the ongoing Larmor precession between the spin states $| \Uparrow \rangle $ and $| \Downarrow \rangle $. In this regime, the cross-correlation oscillates between above- / below-unity values, as the $| \Uparrow \rangle$ state becomes more / less populated than its equilibrium value at long delays when the hole spin is fully depolarized.
We observe remaining oscillations even when no magnetic field was applied indicating a small residual magnetic field of around {$7$~mT}.

%%%%%%% paragraphe supprimé par Pascale%%%%%%%%%%%%%%%%%%%%

%\PS{This analogy is very confusing - I think this is more a internal discussion}\sout{Such experimental results are in strong analogy \nb{NB : reminiscent of ?  je ne comprends pas bien la phrase, le but de le souligner...'strong analogy' me parait trop fort} with the predictions of a related theoretical proposal,  where ground spin states are directly and resonantly read-out, thanks to the spin-dependant optical response of a high-quality optical microcavity \cite{SmirnovReznychenkoAuffevesLanco_PRB}. We note however, that the present version of our experiment does not constitute a direct non-destructive measurement of the ground spin states. Indeed, the detection of a 2nd photon, say, in polarization $R$, is only the signature that the trion state  $| \Uparrow \Downarrow \uparrow \rangle $ has been populated just before photon emission. This indirect read-out has experimental consequences which are discussed below.  }

%\PS{what we really want to report here is the hole spin coherence time, I rewrote this paragraph with the main message put forward - and the discussion avec the hyperfine interaction much reduced.}
 {For further analysis} of the results of Figure 3.c., we compare the experimental data with numerical simulations using the  the same general framework as before. 
To mimic the CW excitation, we consider random excitation events  in the very low excitation regime with only  one excitation event within the relevant delays in the experiment. We obtain  very similar parameters for both the trion lifetime $T_1^{\mathrm{(trion)}}=450 \pm 80$ ps and the hole Land\'e factor $g_h=0.4 \pm 0.01$. Here we can measure the spin coherence time, which reaches $T_2^{\mathrm{(spin)}}=20 \pm 5$ ns for this QD, confirming that  hole spins under LA-assisted excitation are suitable candidates for multi-photon entanglement processes within the LR scheme.
  
\noindent Note that  we  also include the hyperfine interaction of the  electron spin in the trion states that  reduces the fidelity of the optical readout process of the hole spin. The contrast of the oscillations in Fig. 3.c. is well reproduced  by introducing this interaction in the frozen central spin model \cite{Merkulov2002} where  the electron spin is subject to a  randomly-oriented mean field nuclear spin with a standard deviation of $\sigma_{\mathrm{hf}}^{\mathrm{(e)}}=0.8~\mu$eV~\cite{Urbaszek2013}.  Note that this value represents an upper bound since the oscillation contrast is also limited by experimental imperfections such as an imperfect projection to the circular polarization basis.%non-perfect orthogonality of the detection polarization.  

In conclusion, we have shown that acoustic phonon-assisted excitation is an excellent tool to measure a single spin-dynamics in quantum dots opening the path toward the implemention of the Lindner-Rudolph scheme for polarization-encoded cluster state generation. The  detuning of the excitation wavelength compared to the resonant excitation scheme allows to use a simple spectral filtering to remove the excitation laser and  preserve with high fidelity  the optical selection rules. The direct monitoring of the hole spin precession in a low transverse magnetic field also evidenced a heavy-hole spin coherence time of 20 ns in our sample,  a value that is typical for a hole spin in the absence of nuclear spin control~\cite{Warburton2013}. Such coherence time is two orders of magnitude larger than the trion radiative lifetime of a QD inserted in an optical cavity~\cite{Ollivier2020}. We thus anticipate that phonon-assisted excitation scheme should allow the efficient generation of long chains of entangled indistinguishable photons in linear cluster states. %Since acoustic-phonon excitation has also recently been shown to allow for  single-photon generation of near unity indistinguishability and record efficiency for QDs in microcavities\cite{Thomas2021}, we believe that this approach could represent a very direct way to generate linear cluster state at unparalleled rates.

\noindent \textbf{Acknowledgements.} This work was partially supported by the the IAD-ANR support ASTRID program Projet ANR-18-ASTR-0024 LIGHT, the QuantERA ERA-NET Cofund in Quantum Technologies project HIPHOP, the European Union's Horizon 2020 FET OPEN project QLUSTER (Grant ID 862035), the European Union's Horizon 2020 Research and Innovation Programme QUDOT-TECH under the Marie Sklodowska-Curie Grant Agreement No. 861097 and the French RENATECH network, a public grant overseen by the French National Research Agency (ANR) as part of the "Investissements d'Avenir" programme (Labex NanoSaclay, reference: ANR-10-LABX-0035).  N.C. acknowledges support from the Paris Ile-de-France R\'egion in the framework of DIM SIRTEQ.

%\bibliography{Bibliography}
%\bibliography{bibliography-fin}

\begin{thebibliography}{30}%
\makeatletter
\providecommand \@ifxundefined [1]{%
 \@ifx{#1\undefined}
}%
\providecommand \@ifnum [1]{%
 \ifnum #1\expandafter \@firstoftwo
 \else \expandafter \@secondoftwo
 \fi
}%
\providecommand \@ifx [1]{%
 \ifx #1\expandafter \@firstoftwo
 \else \expandafter \@secondoftwo
 \fi
}%
\providecommand \natexlab [1]{#1}%
\providecommand \enquote  [1]{``#1''}%
\providecommand \bibnamefont  [1]{#1}%
\providecommand \bibfnamefont [1]{#1}%
\providecommand \citenamefont [1]{#1}%
\providecommand \href@noop [0]{\@secondoftwo}%
\providecommand \href [0]{\begingroup \@sanitize@url \@href}%
\providecommand \@href[1]{\@@startlink{#1}\@@href}%
\providecommand \@@href[1]{\endgroup#1\@@endlink}%
\providecommand \@sanitize@url [0]{\catcode `\\12\catcode `\$12\catcode
  `\&12\catcode `\#12\catcode `\^12\catcode `\_12\catcode `\%12\relax}%
\providecommand \@@startlink[1]{}%
\providecommand \@@endlink[0]{}%
\providecommand \url  [0]{\begingroup\@sanitize@url \@url }%
\providecommand \@url [1]{\endgroup\@href {#1}{\urlprefix }}%
\providecommand \urlprefix  [0]{URL }%
\providecommand \Eprint [0]{\href }%
\providecommand \doibase [0]{http://dx.doi.org/}%
\providecommand \selectlanguage [0]{\@gobble}%
\providecommand \bibinfo  [0]{\@secondoftwo}%
\providecommand \bibfield  [0]{\@secondoftwo}%
\providecommand \translation [1]{[#1]}%
\providecommand \BibitemOpen [0]{}%
\providecommand \bibitemStop [0]{}%
\providecommand \bibitemNoStop [0]{.\EOS\space}%
\providecommand \EOS [0]{\spacefactor3000\relax}%
\providecommand \BibitemShut  [1]{\csname bibitem#1\endcsname}%
\let\auto@bib@innerbib\@empty
%</preamble>
\bibitem [{\citenamefont {Lindner}\ and\ \citenamefont
  {Rudolph}(2009)}]{lindner_proposal_2009}%
  \BibitemOpen
  \bibfield  {author} {\bibinfo {author} {\bibfnamefont {N.~H.}\ \bibnamefont
  {Lindner}}\ and\ \bibinfo {author} {\bibfnamefont {T.}~\bibnamefont
  {Rudolph}},\ } \bibfield  {title} {\enquote {\bibinfo {title} {Proposal for
  {Pulsed} {On}-{Demand} {Sources} of {Photonic} {Cluster} {State}
  {Strings}},}\ }\href {\doibase 10.1103/PhysRevLett.103.113602} {\bibfield
  {journal} {\bibinfo  {journal} {Physical Review Letters}\ }\textbf {\bibinfo
  {volume} {103}},\ \bibinfo {pages} {113602} (\bibinfo {year}
  {2009})}\BibitemShut {NoStop}%
\bibitem [{\citenamefont {Gao}\ \emph {et~al.}(2015)\citenamefont {Gao},
  \citenamefont {Imamoglu}, \citenamefont {Bernien},\ and\ \citenamefont
  {Hanson}}]{gao_coherent_2015}%
  \BibitemOpen
  \bibfield  {author} {\bibinfo {author} {\bibfnamefont {W.~B.}\ \bibnamefont
  {Gao}}, \bibinfo {author} {\bibfnamefont {A.}~\bibnamefont {Imamoglu}},
  \bibinfo {author} {\bibfnamefont {H.}~\bibnamefont {Bernien}}, \ and\
  \bibinfo {author} {\bibfnamefont {R.}~\bibnamefont {Hanson}},\ }\bibfield
  {title} {\enquote {\bibinfo {title} {Coherent
  manipulation, measurement and entanglement of individual solid-state spins
  using optical fields},}\ }\href {\doibase 10.1038/nphoton.2015.58}
  {\bibfield  {journal} {\bibinfo  {journal} {Nature Photonics}\ }\textbf
  {\bibinfo {volume} {9}},\ \bibinfo {pages} {363--373} (\bibinfo {year}
  {2015})}\BibitemShut {NoStop}%
\bibitem [{\citenamefont {Delteil}\ \emph {et~al.}(2016)\citenamefont
  {Delteil}, \citenamefont {Sun}, \citenamefont {Gao}, \citenamefont {Togan},
  \citenamefont {Faelt},\ and\ \citenamefont
  {Imamoğlu}}]{delteil_generation_2016}%
  \BibitemOpen
  \bibfield  {author} {\bibinfo {author} {\bibfnamefont {A.}~\bibnamefont
  {Delteil}}, \bibinfo {author} {\bibfnamefont {Z.}~\bibnamefont {Sun}},
  \bibinfo {author} {\bibfnamefont {W.-b.}\ \bibnamefont {Gao}}, \bibinfo
  {author} {\bibfnamefont {E.}~\bibnamefont {Togan}}, \bibinfo {author}
  {\bibfnamefont {S.}~\bibnamefont {Faelt}}, \ and\ \bibinfo {author}
  {\bibfnamefont {A.}~\bibnamefont {Imamoğlu}},\ }\bibfield  {title}
  {{\enquote {\bibinfo {title} {Generation of heralded
  entanglement between distant hole spins},}\ }}\href {\doibase
  10.1038/nphys3605} {\bibfield  {journal} {\bibinfo  {journal} {Nature Phys}\
  }\textbf {\bibinfo {volume} {12}},\ \bibinfo {pages} {218--223} (\bibinfo
  {year} {2016})}\BibitemShut {NoStop}%
\bibitem [{\citenamefont {Stockill}\ \emph {et~al.}(2017)\citenamefont
  {Stockill}, \citenamefont {Stanley}, \citenamefont {Huthmacher},
  \citenamefont {Clarke}, \citenamefont {Hugues}, \citenamefont {Miller},
  \citenamefont {Matthiesen}, \citenamefont {Le~Gall},\ and\ \citenamefont
  {Atat\"ure}}]{Stockill2017}%
  \BibitemOpen
  \bibfield  {author} {\bibinfo {author} {\bibfnamefont {R.}~\bibnamefont
  {Stockill}}, \bibinfo {author} {\bibfnamefont {M.~J.}\ \bibnamefont
  {Stanley}}, \bibinfo {author} {\bibfnamefont {L.}~\bibnamefont {Huthmacher}},
  \bibinfo {author} {\bibfnamefont {E.}~\bibnamefont {Clarke}}, \bibinfo
  {author} {\bibfnamefont {M.}~\bibnamefont {Hugues}}, \bibinfo {author}
  {\bibfnamefont {A.~J.}\ \bibnamefont {Miller}}, \bibinfo {author}
  {\bibfnamefont {C.}~\bibnamefont {Matthiesen}}, \bibinfo {author}
  {\bibfnamefont {C.}~\bibnamefont {Le~Gall}}, \ and\ \bibinfo {author}
  {\bibfnamefont {M.}~\bibnamefont {Atat\"ure}},\ }\bibfield  {title} {\enquote
  {\bibinfo {title} {Phase-tuned entangled state generation between distant
  spin qubits},}\ }\href {\doibase 10.1103/PhysRevLett.119.010503} {\bibfield
  {journal} {\bibinfo  {journal} {Phys. Rev. Lett.}\ }\textbf {\bibinfo
  {volume} {119}},\ \bibinfo {pages} {010503} (\bibinfo {year}
  {2017})}\BibitemShut {NoStop}%
\bibitem [{\citenamefont {Bonato}\ \emph {et~al.}(2010)\citenamefont {Bonato},
  \citenamefont {Haupt}, \citenamefont {Oemrawsingh}, \citenamefont {Gudat},
  \citenamefont {Ding}, \citenamefont {van Exter},\ and\ \citenamefont
  {Bouwmeester}}]{Bonato2010}%
  \BibitemOpen
  \bibfield  {author} {\bibinfo {author} {\bibfnamefont {C.}~\bibnamefont
  {Bonato}}, \bibinfo {author} {\bibfnamefont {F.}~\bibnamefont {Haupt}},
  \bibinfo {author} {\bibfnamefont {S.~S.~R.}\ \bibnamefont {Oemrawsingh}},
  \bibinfo {author} {\bibfnamefont {J.}~\bibnamefont {Gudat}}, \bibinfo
  {author} {\bibfnamefont {D.}~\bibnamefont {Ding}}, \bibinfo {author}
  {\bibfnamefont {M.~P.}\ \bibnamefont {van Exter}}, \ and\ \bibinfo {author}
  {\bibfnamefont {D.}~\bibnamefont {Bouwmeester}},\ }\bibfield  {title}
  {\enquote {\bibinfo {title} {Cnot and bell-state analysis in the
  weak-coupling cavity qed regime},}\ }\href {\doibase
  10.1103/PhysRevLett.104.160503} {\bibfield  {journal} {\bibinfo  {journal}
  {Phys. Rev. Lett.}\ }\textbf {\bibinfo {volume} {104}},\ \bibinfo {pages}
  {160503} (\bibinfo {year} {2010})}\BibitemShut {NoStop}%
\bibitem [{\citenamefont {Schwartz}\ \emph {et~al.}(2016)\citenamefont
  {Schwartz}, \citenamefont {Cogan}, \citenamefont {Schmidgall}, \citenamefont
  {Don}, \citenamefont {Gantz}, \citenamefont {Kenneth}, \citenamefont
  {Lindner},\ and\ \citenamefont {Gershoni}}]{schwartz_deterministic_2016}%
  \BibitemOpen
  \bibfield  {author} {\bibinfo {author} {\bibfnamefont {I.}~\bibnamefont
  {Schwartz}}, \bibinfo {author} {\bibfnamefont {D.}~\bibnamefont {Cogan}},
  \bibinfo {author} {\bibfnamefont {E.~R.}\ \bibnamefont {Schmidgall}},
  \bibinfo {author} {\bibfnamefont {Y.}~\bibnamefont {Don}}, \bibinfo {author}
  {\bibfnamefont {L.}~\bibnamefont {Gantz}}, \bibinfo {author} {\bibfnamefont
  {O.}~\bibnamefont {Kenneth}}, \bibinfo {author} {\bibfnamefont {N.~H.}\
  \bibnamefont {Lindner}}, \ and\ \bibinfo {author} {\bibfnamefont
  {D.}~\bibnamefont {Gershoni}},\ }\bibfield  {title} {\enquote {\bibinfo
  {title} {Deterministic {Generation} of a {Cluster} {State} of {Entangled}
  {Photons}},}\ }\href {\doibase 10.1126/science.aah4758} {\bibfield  {journal}
  {\bibinfo  {journal} {Science}\ }\textbf {\bibinfo {volume} {354}},\ \bibinfo
  {pages} {434--437} (\bibinfo {year} {2016})}\BibitemShut {NoStop}%
\bibitem [{\citenamefont {Zwerger}, \citenamefont {D\"ur},\ and\ \citenamefont
  {Briegel}(2012)}]{Zwerger2012}%
  \BibitemOpen
  \bibfield  {author} {\bibinfo {author} {\bibfnamefont {M.}~\bibnamefont
  {Zwerger}}, \bibinfo {author} {\bibfnamefont {W.}~\bibnamefont {D\"ur}}, \
  and\ \bibinfo {author} {\bibfnamefont {H.~J.}\ \bibnamefont {Briegel}},\
  }\bibfield  {title} {\enquote {\bibinfo {title} {Measurement-based quantum
  repeaters},}\ }\href {\doibase 10.1103/PhysRevA.85.062326} {\bibfield
  {journal} {\bibinfo  {journal} {Phys. Rev. A}\ }\textbf {\bibinfo {volume}
  {85}},\ \bibinfo {pages} {062326} (\bibinfo {year} {2012})}\BibitemShut
  {NoStop}%
\bibitem [{\citenamefont {Azuma}, \citenamefont {Tamaki},\ and\ \citenamefont
  {Lo}(2015)}]{Azuma2015}%
  \BibitemOpen
  \bibfield  {author} {\bibinfo {author} {\bibfnamefont {K.}~\bibnamefont
  {Azuma}}, \bibinfo {author} {\bibfnamefont {K.}~\bibnamefont {Tamaki}}, \
  and\ \bibinfo {author} {\bibfnamefont {H.-K.}\ \bibnamefont {Lo}},\
  }\bibfield  {title} {\enquote {\bibinfo {title} {All-photonic quantum
  repeaters},}\ }\href {\doibase 10.1038/ncomms7787} {\bibfield  {journal}
  {\bibinfo  {journal} {Nature Communications}\ }\textbf {\bibinfo {volume}
  {6}},\ \bibinfo {pages} {6787} (\bibinfo {year} {2015})}\BibitemShut
  {NoStop}%
\bibitem [{\citenamefont {Raussendorf}, \citenamefont {Browne},\ and\
  \citenamefont {Briegel}(2003)}]{raussendorf_measurement-based_2003}%
  \BibitemOpen
  \bibfield  {author} {\bibinfo {author} {\bibfnamefont {R.}~\bibnamefont
  {Raussendorf}}, \bibinfo {author} {\bibfnamefont {D.~E.}\ \bibnamefont
  {Browne}}, \ and\ \bibinfo {author} {\bibfnamefont {H.~J.}\ \bibnamefont
  {Briegel}},\ }\bibfield  {title} {\enquote {\bibinfo {title}
  {Measurement-based quantum computation on cluster states},}\ }\href {\doibase
  10.1103/PhysRevA.68.022312} {\bibfield  {journal} {\bibinfo  {journal} {Phys.
  Rev. A}\ }\textbf {\bibinfo {volume} {68}},\ \bibinfo {pages} {022312}
  (\bibinfo {year} {2003})},\ \bibinfo {note} {publisher: American Physical
  Society}\BibitemShut {NoStop}%
\bibitem [{\citenamefont {Somaschi}\ \emph {et~al.}(2016)\citenamefont
  {Somaschi}, \citenamefont {Giesz}, \citenamefont {De~Santis}, \citenamefont
  {Loredo}, \citenamefont {Almeida}, \citenamefont {Hornecker}, \citenamefont
  {Portalupi}, \citenamefont {Grange}, \citenamefont {Antón}, \citenamefont
  {Demory}, \citenamefont {Gómez}, \citenamefont {Sagnes}, \citenamefont
  {Lanzillotti-Kimura}, \citenamefont {Lemaítre}, \citenamefont {Auffeves},
  \citenamefont {White}, \citenamefont {Lanco},\ and\ \citenamefont
  {Senellart}}]{somaschi_near-optimal_2016}%
  \BibitemOpen
  \bibfield  {author} {\bibinfo {author} {\bibfnamefont {N.}~\bibnamefont
  {Somaschi}}, \bibinfo {author} {\bibfnamefont {V.}~\bibnamefont {Giesz}},
  \bibinfo {author} {\bibfnamefont {L.}~\bibnamefont {De~Santis}}, \bibinfo
  {author} {\bibfnamefont {J.~C.}\ \bibnamefont {Loredo}}, \bibinfo {author}
  {\bibfnamefont {M.~P.}\ \bibnamefont {Almeida}}, \bibinfo {author}
  {\bibfnamefont {G.}~\bibnamefont {Hornecker}}, \bibinfo {author}
  {\bibfnamefont {S.~L.}\ \bibnamefont {Portalupi}}, \bibinfo {author}
  {\bibfnamefont {T.}~\bibnamefont {Grange}}, \bibinfo {author} {\bibfnamefont
  {C.}~\bibnamefont {Antón}}, \bibinfo {author} {\bibfnamefont
  {J.}~\bibnamefont {Demory}}, \bibinfo {author} {\bibfnamefont
  {C.}~\bibnamefont {Gómez}}, \bibinfo {author} {\bibfnamefont
  {I.}~\bibnamefont {Sagnes}}, \bibinfo {author} {\bibfnamefont {N.~D.}\
  \bibnamefont {Lanzillotti-Kimura}}, \bibinfo {author} {\bibfnamefont
  {A.}~\bibnamefont {Lemaítre}}, \bibinfo {author} {\bibfnamefont
  {A.}~\bibnamefont {Auffeves}}, \bibinfo {author} {\bibfnamefont {A.~G.}\
  \bibnamefont {White}}, \bibinfo {author} {\bibfnamefont {L.}~\bibnamefont
  {Lanco}}, \ and\ \bibinfo {author} {\bibfnamefont {P.}~\bibnamefont
  {Senellart}},\ }\bibfield  {title} {{\enquote {\bibinfo
  {title} {Near-optimal single-photon sources in the solid state},}\ }}\href
  {\doibase 10.1038/nphoton.2016.23} {\bibfield  {journal} {\bibinfo  {journal}
  {Nature Photon}\ }\textbf {\bibinfo {volume} {10}},\ \bibinfo {pages}
  {340--345} (\bibinfo {year} {2016})}\BibitemShut {NoStop}%
\bibitem [{\citenamefont {Wang}\ \emph {et~al.}(2016)\citenamefont {Wang},
  \citenamefont {Duan}, \citenamefont {Li}, \citenamefont {Chen}, \citenamefont
  {Li}, \citenamefont {He}, \citenamefont {Chen}, \citenamefont {He},
  \citenamefont {Ding}, \citenamefont {Peng}, \citenamefont {Schneider},
  \citenamefont {Kamp}, \citenamefont {H\"ofling}, \citenamefont {Lu},\ and\
  \citenamefont {Pan}}]{Wang2016}%
  \BibitemOpen
  \bibfield  {author} {\bibinfo {author} {\bibfnamefont {H.}~\bibnamefont
  {Wang}}, \bibinfo {author} {\bibfnamefont {Z.-C.}\ \bibnamefont {Duan}},
  \bibinfo {author} {\bibfnamefont {Y.-H.}\ \bibnamefont {Li}}, \bibinfo
  {author} {\bibfnamefont {S.}~\bibnamefont {Chen}}, \bibinfo {author}
  {\bibfnamefont {J.-P.}\ \bibnamefont {Li}}, \bibinfo {author} {\bibfnamefont
  {Y.-M.}\ \bibnamefont {He}}, \bibinfo {author} {\bibfnamefont {M.-C.}\
  \bibnamefont {Chen}}, \bibinfo {author} {\bibfnamefont {Y.}~\bibnamefont
  {He}}, \bibinfo {author} {\bibfnamefont {X.}~\bibnamefont {Ding}}, \bibinfo
  {author} {\bibfnamefont {C.-Z.}\ \bibnamefont {Peng}}, \bibinfo {author}
  {\bibfnamefont {C.}~\bibnamefont {Schneider}}, \bibinfo {author}
  {\bibfnamefont {M.}~\bibnamefont {Kamp}}, \bibinfo {author} {\bibfnamefont
  {S.}~\bibnamefont {H\"ofling}}, \bibinfo {author} {\bibfnamefont {C.-Y.}\
  \bibnamefont {Lu}}, \ and\ \bibinfo {author} {\bibfnamefont {J.-W.}\
  \bibnamefont {Pan}},\ }\bibfield  {title} {\enquote {\bibinfo {title}
  {Near-transform-limited single photons from an efficient solid-state quantum
  emitter},}\ }\href {\doibase 10.1103/PhysRevLett.116.213601} {\bibfield
  {journal} {\bibinfo  {journal} {Phys. Rev. Lett.}\ }\textbf {\bibinfo
  {volume} {116}},\ \bibinfo {pages} {213601} (\bibinfo {year}
  {2016})}\BibitemShut {NoStop}%
\bibitem [{\citenamefont {Rudolph}(2017)}]{Rudolph2017}%
  \BibitemOpen
  \bibfield  {author} {\bibinfo {author} {\bibfnamefont {T.}~\bibnamefont
  {Rudolph}},\ }\bibfield  {title} {\enquote {\bibinfo {title} {Why i am
  optimistic about the silicon-photonic route to quantum computing},}\ }\href
  {\doibase 10.1063/1.4976737} {\bibfield  {journal} {\bibinfo  {journal} {APL
  Photonics}\ }\textbf {\bibinfo {volume} {2}},\ \bibinfo {pages} {030901}
  (\bibinfo {year} {2017})},\ \Eprint
  {http://arxiv.org/abs/https://doi.org/10.1063/1.4976737}
  {https://doi.org/10.1063/1.4976737} \BibitemShut {NoStop}%
\bibitem [{\citenamefont {Cogan}\ \emph {et~al.}(2021)\citenamefont {Cogan},
  \citenamefont {Su}, \citenamefont {Kenneth},\ and\ \citenamefont
  {Gershoni}}]{Cogan2021}%
  \BibitemOpen
  \bibfield  {author} {\bibinfo {author} {\bibfnamefont {D.}~\bibnamefont
  {Cogan}}, \bibinfo {author} {\bibfnamefont {Z.-E.}\ \bibnamefont {Su}},
  \bibinfo {author} {\bibfnamefont {O.}~\bibnamefont {Kenneth}}, \ and\
  \bibinfo {author} {\bibfnamefont {D.}~\bibnamefont {Gershoni}},\ }\bibfield
  {title} {\enquote {\bibinfo {title} {A deterministic source of
  indistinguishable photons in a cluster state},}\ }\href {\doibase
  10.48550/ARXIV.2110.05908} {\  (\bibinfo {year} {2021}),\
  10.48550/ARXIV.2110.05908}\BibitemShut {NoStop}%
\bibitem [{\citenamefont {Kiraz}, \citenamefont {Atat\"ure},\ and\
  \citenamefont {Imamo\ifmmode~\breve{g}\else \u{g}\fi{}lu}(2004)}]{Kiraz2004}%
  \BibitemOpen
  \bibfield  {author} {\bibinfo {author} {\bibfnamefont {A.}~\bibnamefont
  {Kiraz}}, \bibinfo {author} {\bibfnamefont {M.}~\bibnamefont {Atat\"ure}}, \
  and\ \bibinfo {author} {\bibfnamefont {A.}~\bibnamefont
  {Imamo\ifmmode~\breve{g}\else \u{g}\fi{}lu}},\ }\bibfield  {title} {\enquote
  {\bibinfo {title} {Quantum-dot single-photon sources: Prospects for
  applications in linear optics quantum-information processing},}\ }\href
  {\doibase 10.1103/PhysRevA.69.032305} {\bibfield  {journal} {\bibinfo
  {journal} {Phys. Rev. A}\ }\textbf {\bibinfo {volume} {69}},\ \bibinfo
  {pages} {032305} (\bibinfo {year} {2004})}\BibitemShut {NoStop}%
\bibitem [{\citenamefont {He}\ \emph {et~al.}(2013)\citenamefont {He},
  \citenamefont {He}, \citenamefont {Wei}, \citenamefont {Wu}, \citenamefont
  {Atat{\"u}re}, \citenamefont {Schneider}, \citenamefont {H{\"o}fling},
  \citenamefont {Kamp}, \citenamefont {Lu},\ and\ \citenamefont
  {Pan}}]{He2013}%
  \BibitemOpen
  \bibfield  {author} {\bibinfo {author} {\bibfnamefont {Y.-M.}\ \bibnamefont
  {He}}, \bibinfo {author} {\bibfnamefont {Y.}~\bibnamefont {He}}, \bibinfo
  {author} {\bibfnamefont {Y.-J.}\ \bibnamefont {Wei}}, \bibinfo {author}
  {\bibfnamefont {D.}~\bibnamefont {Wu}}, \bibinfo {author} {\bibfnamefont
  {M.}~\bibnamefont {Atat{\"u}re}}, \bibinfo {author} {\bibfnamefont
  {C.}~\bibnamefont {Schneider}}, \bibinfo {author} {\bibfnamefont
  {S.}~\bibnamefont {H{\"o}fling}}, \bibinfo {author} {\bibfnamefont
  {M.}~\bibnamefont {Kamp}}, \bibinfo {author} {\bibfnamefont {C.-Y.}\
  \bibnamefont {Lu}}, \ and\ \bibinfo {author} {\bibfnamefont {J.-W.}\
  \bibnamefont {Pan}},\ }\bibfield  {title} {\enquote {\bibinfo {title}
  {On-demand semiconductor single-photon source with near-unity
  indistinguishability},}\ }\href {\doibase 10.1038/nnano.2012.262} {\bibfield
  {journal} {\bibinfo  {journal} {Nature Nanotechnology}\ }\textbf {\bibinfo
  {volume} {8}},\ \bibinfo {pages} {213--217} (\bibinfo {year}
  {2013})}\BibitemShut {NoStop}%
\bibitem [{\citenamefont {Senellart}, \citenamefont {Solomon},\ and\
  \citenamefont {White}(2017)}]{Senellart2017}%
  \BibitemOpen
  \bibfield  {author} {\bibinfo {author} {\bibfnamefont {P.}~\bibnamefont
  {Senellart}}, \bibinfo {author} {\bibfnamefont {G.}~\bibnamefont {Solomon}},
  \ and\ \bibinfo {author} {\bibfnamefont {A.}~\bibnamefont {White}},\
  }\bibfield  {title} {\enquote {\bibinfo {title} {High-performance
  semiconductor quantum-dot single-photon sources},}\ }\href {\doibase
  10.1038/nnano.2017.218} {\bibfield  {journal} {\bibinfo  {journal} {Nature
  Nanotechnology}\ }\textbf {\bibinfo {volume} {12}},\ \bibinfo {pages}
  {1026--1039} (\bibinfo {year} {2017})}\BibitemShut {NoStop}%
\bibitem [{\citenamefont {Tomm}\ \emph {et~al.}(2021)\citenamefont {Tomm},
  \citenamefont {Javadi}, \citenamefont {Antoniadis}, \citenamefont {Najer},
  \citenamefont {Lobl}, \citenamefont {Korsch}, \citenamefont {Schott},
  \citenamefont {Valentin}, \citenamefont {Wieck}, \citenamefont {Ludwig},\
  and\ \citenamefont {Warburton}}]{Tomm2021}%
  \BibitemOpen
  \bibfield  {author} {\bibinfo {author} {\bibfnamefont {N.}~\bibnamefont
  {Tomm}}, \bibinfo {author} {\bibfnamefont {A.}~\bibnamefont {Javadi}},
  \bibinfo {author} {\bibfnamefont {N.~O.}\ \bibnamefont {Antoniadis}},
  \bibinfo {author} {\bibfnamefont {D.}~\bibnamefont {Najer}}, \bibinfo
  {author} {\bibfnamefont {M.~C.}\ \bibnamefont {Lobl}}, \bibinfo {author}
  {\bibfnamefont {A.~R.}\ \bibnamefont {Korsch}}, \bibinfo {author}
  {\bibfnamefont {R.}~\bibnamefont {Schott}}, \bibinfo {author} {\bibfnamefont
  {S.~R.}\ \bibnamefont {Valentin}}, \bibinfo {author} {\bibfnamefont {A.~D.}\
  \bibnamefont {Wieck}}, \bibinfo {author} {\bibfnamefont {A.}~\bibnamefont
  {Ludwig}}, \ and\ \bibinfo {author} {\bibfnamefont {R.~J.}\ \bibnamefont
  {Warburton}},\ }\bibfield  {title} {\enquote {\bibinfo {title} {A bright and
  fast source of coherent single photons},}\ }\href {\doibase
  10.1038/s41565-020-00831-x} {\bibfield  {journal} {\bibinfo  {journal}
  {Nature Nanotechnology}\ } (\bibinfo {year} {2021}),\
  10.1038/s41565-020-00831-x}\BibitemShut {NoStop}%
\bibitem [{\citenamefont {Tiurev}\ \emph {et~al.}(2021)\citenamefont {Tiurev},
  \citenamefont {Mirambell}, \citenamefont {Lauritzen}, \citenamefont {Appel},
  \citenamefont {Tiranov}, \citenamefont {Lodahl},\ and\ \citenamefont
  {S\o{}rensen}}]{Tiurev2021}%
  \BibitemOpen
  \bibfield  {author} {\bibinfo {author} {\bibfnamefont {K.}~\bibnamefont
  {Tiurev}}, \bibinfo {author} {\bibfnamefont {P.~L.}\ \bibnamefont
  {Mirambell}}, \bibinfo {author} {\bibfnamefont {M.~B.}\ \bibnamefont
  {Lauritzen}}, \bibinfo {author} {\bibfnamefont {M.~H.}\ \bibnamefont
  {Appel}}, \bibinfo {author} {\bibfnamefont {A.}~\bibnamefont {Tiranov}},
  \bibinfo {author} {\bibfnamefont {P.}~\bibnamefont {Lodahl}}, \ and\ \bibinfo
  {author} {\bibfnamefont {A.~S.}\ \bibnamefont {S\o{}rensen}},\ }\bibfield
  {title} {\enquote {\bibinfo {title} {Fidelity of time-bin-entangled
  multiphoton states from a quantum emitter},}\ }\href {\doibase
  10.1103/PhysRevA.104.052604} {\bibfield  {journal} {\bibinfo  {journal}
  {Phys. Rev. A}\ }\textbf {\bibinfo {volume} {104}},\ \bibinfo {pages}
  {052604} (\bibinfo {year} {2021})}\BibitemShut {NoStop}%
\bibitem [{\citenamefont {Tiurev}\ \emph {et~al.}(2022)\citenamefont {Tiurev},
  \citenamefont {Appel}, \citenamefont {Mirambell}, \citenamefont {Lauritzen},
  \citenamefont {Tiranov}, \citenamefont {Lodahl},\ and\ \citenamefont
  {S\o{}rensen}}]{Tiurev2022}%
  \BibitemOpen
  \bibfield  {author} {\bibinfo {author} {\bibfnamefont {K.}~\bibnamefont
  {Tiurev}}, \bibinfo {author} {\bibfnamefont {M.~H.}\ \bibnamefont {Appel}},
  \bibinfo {author} {\bibfnamefont {P.~L.}\ \bibnamefont {Mirambell}}, \bibinfo
  {author} {\bibfnamefont {M.~B.}\ \bibnamefont {Lauritzen}}, \bibinfo {author}
  {\bibfnamefont {A.}~\bibnamefont {Tiranov}}, \bibinfo {author} {\bibfnamefont
  {P.}~\bibnamefont {Lodahl}}, \ and\ \bibinfo {author} {\bibfnamefont {A.~S.}\
  \bibnamefont {S\o{}rensen}},\ }\bibfield  {title} {\enquote {\bibinfo {title}
  {High-fidelity multiphoton-entangled cluster state with solid-state quantum
  emitters in photonic nanostructures},}\ }\href {\doibase
  10.1103/PhysRevA.105.L030601} {\bibfield  {journal} {\bibinfo  {journal}
  {Phys. Rev. A}\ }\textbf {\bibinfo {volume} {105}},\ \bibinfo {pages}
  {L030601} (\bibinfo {year} {2022})}\BibitemShut {NoStop}%
\bibitem [{\citenamefont {Lee}\ \emph {et~al.}(2019)\citenamefont {Lee},
  \citenamefont {Villa}, \citenamefont {Bennett}, \citenamefont {Stevenson},
  \citenamefont {Ellis}, \citenamefont {Farrer}, \citenamefont {Ritchie},\ and\
  \citenamefont {Shields}}]{Lee2019}%
  \BibitemOpen
  \bibfield  {author} {\bibinfo {author} {\bibfnamefont {J.~P.}\ \bibnamefont
  {Lee}}, \bibinfo {author} {\bibfnamefont {B.}~\bibnamefont {Villa}}, \bibinfo
  {author} {\bibfnamefont {A.~J.}\ \bibnamefont {Bennett}}, \bibinfo {author}
  {\bibfnamefont {R.~M.}\ \bibnamefont {Stevenson}}, \bibinfo {author}
  {\bibfnamefont {D.~J.~P.}\ \bibnamefont {Ellis}}, \bibinfo {author}
  {\bibfnamefont {I.}~\bibnamefont {Farrer}}, \bibinfo {author} {\bibfnamefont
  {D.~A.}\ \bibnamefont {Ritchie}}, \ and\ \bibinfo {author} {\bibfnamefont
  {A.~J.}\ \bibnamefont {Shields}},\ }\bibfield  {title} {\enquote {\bibinfo
  {title} {A quantum dot as a source of time-bin entangled multi-photon
  states},}\ }\href {\doibase 10.1088/2058-9565/ab0a9b} {\bibfield  {journal}
  {\bibinfo  {journal} {Quantum Science and Technology}\ }\textbf {\bibinfo
  {volume} {4}},\ \bibinfo {pages} {025011} (\bibinfo {year}
  {2019})}\BibitemShut {NoStop}%
\bibitem [{\citenamefont {Ardelt}\ \emph {et~al.}(2014)\citenamefont {Ardelt},
  \citenamefont {Hanschke}, \citenamefont {Fischer}, \citenamefont {M\"uller},
  \citenamefont {Kleinkauf}, \citenamefont {Koller}, \citenamefont {Bechtold},
  \citenamefont {Simmet}, \citenamefont {Wierzbowski}, \citenamefont {Riedl},
  \citenamefont {Abstreiter},\ and\ \citenamefont {Finley}}]{Finley2014}%
  \BibitemOpen
  \bibfield  {author} {\bibinfo {author} {\bibfnamefont {P.-L.}\ \bibnamefont
  {Ardelt}}, \bibinfo {author} {\bibfnamefont {L.}~\bibnamefont {Hanschke}},
  \bibinfo {author} {\bibfnamefont {K.~A.}\ \bibnamefont {Fischer}}, \bibinfo
  {author} {\bibfnamefont {K.}~\bibnamefont {M\"uller}}, \bibinfo {author}
  {\bibfnamefont {A.}~\bibnamefont {Kleinkauf}}, \bibinfo {author}
  {\bibfnamefont {M.}~\bibnamefont {Koller}}, \bibinfo {author} {\bibfnamefont
  {A.}~\bibnamefont {Bechtold}}, \bibinfo {author} {\bibfnamefont
  {T.}~\bibnamefont {Simmet}}, \bibinfo {author} {\bibfnamefont
  {J.}~\bibnamefont {Wierzbowski}}, \bibinfo {author} {\bibfnamefont
  {H.}~\bibnamefont {Riedl}}, \bibinfo {author} {\bibfnamefont
  {G.}~\bibnamefont {Abstreiter}}, \ and\ \bibinfo {author} {\bibfnamefont
  {J.~J.}\ \bibnamefont {Finley}},\ }\bibfield  {title} {\enquote {\bibinfo
  {title} {Dissipative preparation of the exciton and biexciton in
  self-assembled quantum dots on picosecond time scales},}\ }\href {\doibase
  10.1103/PhysRevB.90.241404} {\bibfield  {journal} {\bibinfo  {journal} {Phys.
  Rev. B}\ }\textbf {\bibinfo {volume} {90}},\ \bibinfo {pages} {241404}
  (\bibinfo {year} {2014})}\BibitemShut {NoStop}%
\bibitem [{\citenamefont {Cosacchi}\ \emph {et~al.}(2019)\citenamefont
  {Cosacchi}, \citenamefont {Ungar}, \citenamefont {Cygorek}, \citenamefont
  {Vagov},\ and\ \citenamefont {Axt}}]{Cosacchi2019}%
  \BibitemOpen
  \bibfield  {author} {\bibinfo {author} {\bibfnamefont {M.}~\bibnamefont
  {Cosacchi}}, \bibinfo {author} {\bibfnamefont {F.}~\bibnamefont {Ungar}},
  \bibinfo {author} {\bibfnamefont {M.}~\bibnamefont {Cygorek}}, \bibinfo
  {author} {\bibfnamefont {A.}~\bibnamefont {Vagov}}, \ and\ \bibinfo {author}
  {\bibfnamefont {V.~M.}\ \bibnamefont {Axt}},\ }\bibfield  {title} {\enquote
  {\bibinfo {title} {Emission-frequency separated high quality single-photon
  sources enabled by phonons},}\ }\href {\doibase
  10.1103/PhysRevLett.123.017403} {\bibfield  {journal} {\bibinfo  {journal}
  {Phys. Rev. Lett.}\ }\textbf {\bibinfo {volume} {123}},\ \bibinfo {pages}
  {017403} (\bibinfo {year} {2019})}\BibitemShut {NoStop}%
\bibitem [{\citenamefont {Gustin}\ and\ \citenamefont
  {Hughes}(2020)}]{GustinHughes2020}%
  \BibitemOpen
  \bibfield  {author} {\bibinfo {author} {\bibfnamefont {C.}~\bibnamefont
  {Gustin}}\ and\ \bibinfo {author} {\bibfnamefont {S.}~\bibnamefont
  {Hughes}},\ }\bibfield  {title} {\enquote {\bibinfo {title} {Efficient
  pulse-excitation techniques for single photon sources from quantum dots in
  optical cavities},}\ }\href {\doibase 10.1002/qute.201900073} {\bibfield
  {journal} {\bibinfo  {journal} {Advanced Quantum Technologies}\ }\textbf
  {\bibinfo {volume} {3}},\ \bibinfo {pages} {1900073} (\bibinfo {year}
  {2020})}\BibitemShut {NoStop}%
\bibitem [{\citenamefont {Thomas}\ \emph {et~al.}(2021)\citenamefont {Thomas},
  \citenamefont {Billard}, \citenamefont {Coste}, \citenamefont {Wein},
  \citenamefont {{Priya}}, \citenamefont {Ollivier}, \citenamefont {Krebs},
  \citenamefont {Tazaïrt}, \citenamefont {Harouri}, \citenamefont {Lemaitre},
  \citenamefont {Sagnes}, \citenamefont {Anton}, \citenamefont {Lanco},
  \citenamefont {Somaschi}, \citenamefont {Loredo},\ and\ \citenamefont
  {Senellart}}]{thomas_bright_2021}%
  \BibitemOpen
  \bibfield  {author} {\bibinfo {author} {\bibfnamefont {S.}~\bibnamefont
  {Thomas}}, \bibinfo {author} {\bibfnamefont {M.}~\bibnamefont {Billard}},
  \bibinfo {author} {\bibfnamefont {N.}~\bibnamefont {Coste}}, \bibinfo
  {author} {\bibfnamefont {S.}~\bibnamefont {Wein}}, \bibinfo {author}
  {\bibnamefont {{Priya}}}, \bibinfo {author} {\bibfnamefont {H.}~\bibnamefont
  {Ollivier}}, \bibinfo {author} {\bibfnamefont {O.}~\bibnamefont {Krebs}},
  \bibinfo {author} {\bibfnamefont {L.}~\bibnamefont {Tazaïrt}}, \bibinfo
  {author} {\bibfnamefont {A.}~\bibnamefont {Harouri}}, \bibinfo {author}
  {\bibfnamefont {A.}~\bibnamefont {Lemaitre}}, \bibinfo {author}
  {\bibfnamefont {I.}~\bibnamefont {Sagnes}}, \bibinfo {author} {\bibfnamefont
  {C.}~\bibnamefont {Anton}}, \bibinfo {author} {\bibfnamefont
  {L.}~\bibnamefont {Lanco}}, \bibinfo {author} {\bibfnamefont
  {N.}~\bibnamefont {Somaschi}}, \bibinfo {author} {\bibfnamefont
  {J.}~\bibnamefont {Loredo}}, \ and\ \bibinfo {author} {\bibfnamefont
  {P.}~\bibnamefont {Senellart}},\ }\bibfield  {title} {{ {\bibinfo {title} {Bright {Polarized} {Single}-{Photon} {Source}
  {Based} on a {Linear} {Dipole}},}\ }}\href {\doibase
  10.1103/PhysRevLett.126.233601} {\bibfield  {journal} {\bibinfo  {journal}
  {Phys. Rev. Lett.}\ }\textbf {\bibinfo {volume} {126}},\ \bibinfo {pages}
  {233601} (\bibinfo {year} {2021})}\BibitemShut {NoStop}%
\bibitem [{\citenamefont {Reindl}\ \emph {et~al.}(2019)\citenamefont {Reindl},
  \citenamefont {Weber}, \citenamefont {Huber}, \citenamefont {Schimpf},
  \citenamefont {Covre~da Silva}, \citenamefont {Portalupi}, \citenamefont
  {Trotta}, \citenamefont {Michler},\ and\ \citenamefont
  {Rastelli}}]{Rastelli2019}%
  \BibitemOpen
  \bibfield  {author} {\bibinfo {author} {\bibfnamefont {M.}~\bibnamefont
  {Reindl}}, \bibinfo {author} {\bibfnamefont {J.~H.}\ \bibnamefont {Weber}},
  \bibinfo {author} {\bibfnamefont {D.}~\bibnamefont {Huber}}, \bibinfo
  {author} {\bibfnamefont {C.}~\bibnamefont {Schimpf}}, \bibinfo {author}
  {\bibfnamefont {S.~F.}\ \bibnamefont {Covre~da Silva}}, \bibinfo {author}
  {\bibfnamefont {S.~L.}\ \bibnamefont {Portalupi}}, \bibinfo {author}
  {\bibfnamefont {R.}~\bibnamefont {Trotta}}, \bibinfo {author} {\bibfnamefont
  {P.}~\bibnamefont {Michler}}, \ and\ \bibinfo {author} {\bibfnamefont
  {A.}~\bibnamefont {Rastelli}},\ }\bibfield  {title} {\enquote {\bibinfo
  {title} {Highly indistinguishable single photons from incoherently excited
  quantum dots},}\ }\href {\doibase 10.1103/PhysRevB.100.155420} {\bibfield
  {journal} {\bibinfo  {journal} {Phys. Rev. B}\ }\textbf {\bibinfo {volume}
  {100}},\ \bibinfo {pages} {155420} (\bibinfo {year} {2019})}\BibitemShut
  {NoStop}%
\bibitem [{\citenamefont {Trifonov}\ \emph {et~al.}(2021)\citenamefont
  {Trifonov}, \citenamefont {Akimov}, \citenamefont {Golub}, \citenamefont
  {Ivchenko}, \citenamefont {Yugova}, \citenamefont {Kosarev}, \citenamefont
  {Scholz}, \citenamefont {Sgroi}, \citenamefont {Ludwig}, \citenamefont
  {Wieck}, \citenamefont {Yakovlev},\ and\ \citenamefont
  {Bayer}}]{Trifonov2021}%
  \BibitemOpen
  \bibfield  {author} {\bibinfo {author} {\bibfnamefont {A.~V.}\ \bibnamefont
  {Trifonov}}, \bibinfo {author} {\bibfnamefont {I.~A.}\ \bibnamefont
  {Akimov}}, \bibinfo {author} {\bibfnamefont {L.~E.}\ \bibnamefont {Golub}},
  \bibinfo {author} {\bibfnamefont {E.~L.}\ \bibnamefont {Ivchenko}}, \bibinfo
  {author} {\bibfnamefont {I.~A.}\ \bibnamefont {Yugova}}, \bibinfo {author}
  {\bibfnamefont {A.~N.}\ \bibnamefont {Kosarev}}, \bibinfo {author}
  {\bibfnamefont {S.~E.}\ \bibnamefont {Scholz}}, \bibinfo {author}
  {\bibfnamefont {C.}~\bibnamefont {Sgroi}}, \bibinfo {author} {\bibfnamefont
  {A.}~\bibnamefont {Ludwig}}, \bibinfo {author} {\bibfnamefont {A.~D.}\
  \bibnamefont {Wieck}}, \bibinfo {author} {\bibfnamefont {D.~R.}\ \bibnamefont
  {Yakovlev}}, \ and\ \bibinfo {author} {\bibfnamefont {M.}~\bibnamefont
  {Bayer}},\ }\bibfield  {title} {\enquote {\bibinfo {title} {Strong
  enhancement of heavy-hole landé factor $q$ in ingaas symmetric quantum dots
  revealed by coherent optical spectroscopy},}\ }\href {\doibase
  10.48550/ARXIV.2103.13653} {\  (\bibinfo {year} {2021}),\
  10.48550/ARXIV.2103.13653}\BibitemShut {NoStop}%
\bibitem [{\citenamefont {Urbaszek}\ \emph {et~al.}(2013)\citenamefont
  {Urbaszek}, \citenamefont {Marie}, \citenamefont {Amand}, \citenamefont
  {Krebs}, \citenamefont {Voisin}, \citenamefont {Maletinsky}, \citenamefont
  {H\"ogele},\ and\ \citenamefont {Imamoglu}}]{Urbaszek2013}%
  \BibitemOpen
  \bibfield  {author} {\bibinfo {author} {\bibfnamefont {B.}~\bibnamefont
  {Urbaszek}}, \bibinfo {author} {\bibfnamefont {X.}~\bibnamefont {Marie}},
  \bibinfo {author} {\bibfnamefont {T.}~\bibnamefont {Amand}}, \bibinfo
  {author} {\bibfnamefont {O.}~\bibnamefont {Krebs}}, \bibinfo {author}
  {\bibfnamefont {P.}~\bibnamefont {Voisin}}, \bibinfo {author} {\bibfnamefont
  {P.}~\bibnamefont {Maletinsky}}, \bibinfo {author} {\bibfnamefont
  {A.}~\bibnamefont {H\"ogele}}, \ and\ \bibinfo {author} {\bibfnamefont
  {A.}~\bibnamefont {Imamoglu}},\ }\bibfield  {title} {\enquote {\bibinfo
  {title} {Nuclear spin physics in quantum dots: An optical investigation},}\
  }\href {\doibase 10.1103/RevModPhys.85.79} {\bibfield  {journal} {\bibinfo
  {journal} {Rev. Mod. Phys.}\ }\textbf {\bibinfo {volume} {85}},\ \bibinfo
  {pages} {79--133} (\bibinfo {year} {2013})}\BibitemShut {NoStop}%
\bibitem [{\citenamefont {Merkulov}, \citenamefont {Efros},\ and\ \citenamefont
  {Rosen}(2002)}]{Merkulov2002}%
  \BibitemOpen
  \bibfield  {author} {\bibinfo {author} {\bibfnamefont {I.~A.}\ \bibnamefont
  {Merkulov}}, \bibinfo {author} {\bibfnamefont {A.~L.}\ \bibnamefont {Efros}},
  \ and\ \bibinfo {author} {\bibfnamefont {M.}~\bibnamefont {Rosen}},\
  }\bibfield  {title} {\enquote {\bibinfo {title} {Electron spin relaxation by
  nuclei in semiconductor quantum dots},}\ }\href {\doibase
  10.1103/PhysRevB.65.205309} {\bibfield  {journal} {\bibinfo  {journal} {Phys.
  Rev. B}\ }\textbf {\bibinfo {volume} {65}},\ \bibinfo {pages} {205309}
  (\bibinfo {year} {2002})}\BibitemShut {NoStop}%
\bibitem [{\citenamefont {Warburton}(2013)}]{Warburton2013}%
  \BibitemOpen
  \bibfield  {author} {\bibinfo {author} {\bibfnamefont {R.~J.}\ \bibnamefont
  {Warburton}},\ }\bibfield  {title} {\enquote {\bibinfo {title} {Single spins
  in self-assembled quantum dots},}\ }\href@noop {} {\bibfield  {journal}
  {\bibinfo  {journal} {Nature Materials}\ }\textbf {\bibinfo {volume} {12}},\
  \bibinfo {pages} {483--493} (\bibinfo {year} {2013})}\BibitemShut {NoStop}%
\bibitem [{\citenamefont {Ollivier}\ \emph {et~al.}(2020)\citenamefont
  {Ollivier}, \citenamefont {Maillette~de Buy~Wenniger}, \citenamefont
  {Thomas}, \citenamefont {Wein}, \citenamefont {Harouri}, \citenamefont
  {Coppola}, \citenamefont {Hilaire}, \citenamefont {Millet}, \citenamefont
  {Lema{\'i}tre}, \citenamefont {Sagnes}, \citenamefont {Krebs}, \citenamefont
  {Lanco}, \citenamefont {Loredo}, \citenamefont {Ant{\'o}n}, \citenamefont
  {Somaschi},\ and\ \citenamefont {Senellart}}]{Ollivier2020}%
  \BibitemOpen
  \bibfield  {author} {\bibinfo {author} {\bibfnamefont {H.}~\bibnamefont
  {Ollivier}}, \bibinfo {author} {\bibfnamefont {I.}~\bibnamefont {Maillette~de
  Buy~Wenniger}}, \bibinfo {author} {\bibfnamefont {S.}~\bibnamefont {Thomas}},
  \bibinfo {author} {\bibfnamefont {S.~C.}\ \bibnamefont {Wein}}, \bibinfo
  {author} {\bibfnamefont {A.}~\bibnamefont {Harouri}}, \bibinfo {author}
  {\bibfnamefont {G.}~\bibnamefont {Coppola}}, \bibinfo {author} {\bibfnamefont
  {P.}~\bibnamefont {Hilaire}}, \bibinfo {author} {\bibfnamefont
  {C.}~\bibnamefont {Millet}}, \bibinfo {author} {\bibfnamefont
  {A.}~\bibnamefont {Lema{\'i}tre}}, \bibinfo {author} {\bibfnamefont
  {I.}~\bibnamefont {Sagnes}}, \bibinfo {author} {\bibfnamefont
  {O.}~\bibnamefont {Krebs}}, \bibinfo {author} {\bibfnamefont
  {L.}~\bibnamefont {Lanco}}, \bibinfo {author} {\bibfnamefont {J.~C.}\
  \bibnamefont {Loredo}}, \bibinfo {author} {\bibfnamefont {C.}~\bibnamefont
  {Ant{\'o}n}}, \bibinfo {author} {\bibfnamefont {N.}~\bibnamefont {Somaschi}},
  \ and\ \bibinfo {author} {\bibfnamefont {P.}~\bibnamefont {Senellart}},\
  }\bibfield  {title} {\enquote {\bibinfo {title} {Reproducibility of
  high-performance quantum dot single-photon sources},}\ }\href {\doibase
  10.1021/acsphotonics.9b01805} {\bibfield  {journal} {\bibinfo  {journal} {ACS
  Photonics}\ }\textbf {\bibinfo {volume} {7}},\ \bibinfo {pages} {1050--1059}
  (\bibinfo {year} {2020})}\BibitemShut {NoStop}%
\end{thebibliography}

\end{document}